# AMPLIFICATION OF ACOUSTIC WAVE IN GaN IN THE PRESENCE OF SLOWLY CHANGING PERIODIC ELECTRIC FIELD

N. G. Mensah

*Department of Mathematics and Statistics, University of Cape Coast, Ghana*

**Abstract:**

Acoustic wave propagation in bulk GaN Semiconductor in the presence of a slowly changing a.c electric field and a constant electric field has been studied. Analytical expression for the attenuation (amplification) coefficient has been obtained. It is shown that the a.c electric field is acting as a modulator and that the acoustoelectric gain increases to a maximum value and then fall off as the acoustic wave frequency is increased. Also for a a.c electric field of $4 \times 10^4$ V/CM the acoustoelectric gain remains zero until the acoustic wave frequency is about $5.8 \times 10^{12}$ Hz then a resonant amplification peak appears. If the a.c electric field is increased a second resonant peak accurs at $8.7 \times 10^{12}$ Hz. It is therefore suggested that the sample can be used as a maser.

**INTRODUCTION**

The measure of the attenuation of acoustic wave permit one to study the influence on the propagation behavior of any such property of the solid that is sufficiently well coupled to
the lattice; for example Electron – phonon interaction [1 – 5], Thermo-elastic or heating effect [6 – 8], Magneto-elastic loss effect in ferromagnetic material [9 – 10], Phonon-phonon interaction [1,5], Acoustoelectric effect [12 – 16], Acoustomagneto-electric effect [17-22], acoustothermal effect [23].

As a result of the aforementioned reasons the study of attenuation of acoustic waves has received a lot of attention [24 – 29]. This phenomenon can be explained as follows:

When acoustic wave passes through a material it may interact with various elementary excitations. In such an interaction the acoustic wave may lose energy or gain energy under certain circumstances. The latter is known as amplification and the former as attenuation of acoustic wave. The idea of acoustic wave amplification was theoretically predicted in 1956 by Tolpygo and Uritskii [30] and by Weinreich [24] and experimentally observed in CdS by Hatson et al [25] and in n-Ge by Pomerant [31]. More recent works conducted on superlattice material revealed a lot of interesting results [32]. Work has also been done on a two-dimensional system by Govorov et al [29] and on GaN [27, 28].

This paper is aimed at studying how the attenuation amplification is effected by a slowly changing electric field on the drift velocity of the electrons in bulk GaN. Currently, the Ga N related material systems are receiving much attention because of present and potential devices applications, such as blue lasers, UV detectors and high power high temperature field effect transistors. It is our opinion that it will also find important application in acoustoelectronics.

2. MACROSCOPIC THEORY

In discussing the interaction of acoustic waves with mobile carriers, two limiting cases, $ql \ll 1$ and $ql \gg 1$ are conveniently considered ($q = \frac{2\pi}{\lambda}$, where $\lambda$ acoustic wave length and $l$ is the mean free path of the carriers). The acoustic wavelength $\lambda$ is usually large compared with $l$ at ultrasonic frequencies so that $ql$ is considerably less than unity. This condition allows the treatment of the acoustic wave as a classical wave which perturbs the carrier distribution, and hence makes the theory macroscopic in nature.

Consider a plane acoustic wave, either pure shear or pure longitudinal, propagating along the x – direction of an orthogonal set of co-ordinates. If u is the particle displacement along x (longitudinal wave) or in the y,z plane (shear waves), then the strain S is $\frac{\partial u}{\partial x}$. As a result, only longitudinal piezoelectric fields need to be considered. Thus reducing the problem essentially to one-dimensional case.

Using the electromechanical equations of a piezoelectric semiconductor, the displacement D is given by the equation

$$D = e_p S + \varepsilon E \qquad (1)$$

and the stress T has the form

$$T = cS - e_p E \qquad (2)$$

where c is the elastic constant at constant applied electric field E and $e_p$ is piezoelectric constant. S and $\varepsilon$ are the strain and the dielectric constant at constant strain.

For the piezoelectric equations, the wave equation in an elastic medium can be written as

$$\frac{\partial T}{\partial X} = \rho \frac{\partial^2 u}{\partial t^2} = c \frac{\partial^2 u}{\partial x^2} - e_p \frac{\partial E}{\partial X} \qquad (3)$$

where $\rho$ is the lattice density. The current density is given as

$$j = e(\eta_0 + n)\mu E + eD_n \frac{\partial n}{\partial x} \qquad (4)$$

where $\mu$ is the electron, mobility, $D_n$ is the diffusion coefficient, $n_0$ is the equilibrium density of conduction electron and n is the variation caused by the acoustic wave.

Eq(4) coupled with the Poisson and continuity equations i.e.

$$\frac{\partial D}{\partial X} = -en \qquad (5)$$

$$\frac{\partial j}{\partial x} = e \frac{\partial n}{dt} \qquad (6)$$

lead to material displacement $D$ and electric field $E$, equations

$$E = E_0 + E_1 \exp[i(k_s x - \omega_0 t)] \qquad (7)$$

$$D = D_0 + D_1 \exp[i(k_s x - \omega_0 t)] \qquad (8)$$

The $E_0$ in Eq(7) represents an external dc electric field applied to produce a steady electric drift. The linear small-signal theory implies that the term $\eta.E_1$ in Eq(4) is neglected. Combining Eq(4) to Eq(6) and eliminating E from Eq(4) after cumbersome manipulation gives a new relation

$$\rho \frac{\partial^2 u}{\partial t^2} = c' \frac{\partial^2 u}{\partial x^2} \quad (9)$$

where

$$C' = C\left[1 + \frac{e_p^2}{c\varepsilon}\left\{\frac{\left[1+\left(\frac{\omega_c}{\gamma^2\omega_0^2}\right)\left(\frac{\omega_0^2}{\omega_c\omega}\right)\left[1+\frac{\omega_0^2}{\omega_c\omega_D}\right]-i\left(\frac{\omega_c}{\gamma\omega}\right)\right]}{\left[1+\left(\frac{\omega_c}{\gamma^2\omega_0^2}\right)\left[1+\frac{\omega_0^2}{\omega_c\omega_D}\right]\right]^2}\right\}\right] \quad (10)$$

here $\omega_D = \frac{V_0^2}{D_n}$ is the electron diffusion frequency, $\gamma = 1 - \frac{\mu E}{V_0} = 1 - \frac{V_d}{V_0}$ is the drift parameter and $\omega_c = \frac{\eta_0 e\mu}{\varepsilon} = \frac{\delta}{\varepsilon}$ is the conduction relaxation frequency. Note that $V_0 = \left(\frac{c}{\rho}\right)^{1/2}$ is the speed of sound in material. The solution for Eq(9) can now be sought as

$$u = u_1 \exp[i(k_s x - \omega_0 t)] \quad (11)$$

here the propagation constant $k_s = \frac{\omega_0}{V_0} + i\alpha$ ; $\alpha$ being the attenuation coefficient.

The attenuation or absorption coefficient $\alpha$ and the acoustic wave velocity $V_s$ are then found as

$$\alpha = \omega_0 \rho^{1/2} I_m\left((c')^{-1/2}\right) \quad (12)$$

$$V_s = \rho^{-1/2} R_e\left[(c')^{1/2}\right] \quad (13)$$

hence

$$\alpha = \omega_0 \left(\frac{\rho}{c}\right)^{1/2}\left(\frac{e_p^2}{2c\varepsilon}\right)\left[\frac{\frac{\omega_c}{\gamma\omega_0}}{1+\left(\frac{\omega_c^2}{\gamma^2\omega_0^2}\right)\left[1+\frac{\omega_0^2}{\omega_c\omega_D}\right]}\right] \quad (14)$$

This equation can be written in terms of electromechanical coupling constant as

$$\alpha = \frac{K^2}{2}\frac{\omega_c}{v_0\gamma}\left[1+\left(\frac{\omega_c^2}{\gamma^2\omega_0^2}\right)\left[1+\frac{\omega_0^2}{\omega_c\omega_D}\right]\right]^{-1} \quad (15)$$

where $K = \ell\rho(\varepsilon c)^{1/2}$

is the electromechanical coupling constant which is a measure for the strength of piezoelectricity.

## DETERMINATION OF $\alpha$ IN THE PRESENCE OF CHANGING PERIODIC ELECTRIC FIELD $E(t)$

Consider acoustic wave moving through a sample only once at $x=0$ (neglecting reflection at the edges) under the influence of changing periodic field $E(t)$ with frequency $\omega$ satisfying the relation $2\pi\frac{V_0}{L} \leq \omega$ ( $L$ is the length of the sample). The intensity of acoustic wave $I(x,t)$ will satisfy the following equation

$$\frac{\partial I}{\partial t} + V_0\frac{\partial I}{\partial X} = -\alpha(t)V_0 I \quad (16)$$

where $\alpha(t)$ is the coefficient of attenuation (amplification) of sound at time $t$.

The boundary condition is given as

$$I(x,t)\big|_{x=0} = I_0 \quad (17)$$

It is easy to show that the solution of Eq(16) with Eq(17) as the boundary condition $X = L$ is given as

$$I(L,t) = I_0 \exp\left[-V_0 \int_{t-\frac{L}{V_0}}^{t} \alpha(\tau)d\tau\right] \quad (18)$$

Eq(18) represent the intensity of sound at the end of the sample. Lets assume the flight time taken by sound to move through the sample to be $t = n\frac{2\pi}{\omega} + \theta$. Where $n = 1, 2, 3,...$ and $\omega$ is the frequency of the external electric field. Note that $\omega$ satisfy the condition $2\pi\frac{V_0}{L} \leq \omega$ and $\theta$ satisfy $0 < \theta < \frac{2\pi}{\omega}$. Then solving the integral in the exponent for $t \gg \frac{L}{V_O}$ we obtain

$$I(L,t) = I_0 \exp\{-\overline{\alpha}L + [\overline{\alpha} - \overline{\alpha}_t(t)]V_0\theta\} \quad (19)$$

where

$$\overline{\alpha} = \frac{\omega}{2\pi} \int_0^{\frac{2\pi}{\omega}} \alpha(\tau)d\tau \quad (20)$$

is the average of the attenuation (amplification) coefficient over the period of the external field. And

$$\Delta_\theta \overline{\alpha}(t) = \frac{1}{\theta} \int_0^\theta \alpha\left(t' + t - \frac{l}{V_0}\right)dt' \quad (21)$$

if $\frac{L}{V_0} \eta T$ then the acoustic wave intensity at the end of the sample becomes

$$I(L) = L_0 \exp(-\overline{\alpha}L) \quad (22)$$

Hence the dependence of $\alpha$ on the external periodic field at the end of the sample becomes constant and is equal to $\overline{\alpha}$.

For $\Delta \neq 0$ maximum amplitude relative to the fluctuation at the end of the sample becomes

$$\left|\frac{\Delta I}{I}\right| < |1 - \exp(\alpha_{max} - \alpha_{min})| \quad (23)$$

for sufficiently small period T such that

$$|(\alpha_{max} - \alpha_{min})V_0 T| \ll 1 \quad (24)$$

The intensity of the acoustic wave at the end of the sample becomes almost constant and is given by the Eq(22).

For a slowly changing external field (i.e. $\omega\tau \ll 1$) $E(t) = E_0 + E_1 \sin(\omega t)$ we replace the constant electric field E in Eq(14) with $E_0 + E_1 \sin(\omega t)$ and average the result over the period as indicated in Eq(20). The following expression is then obtained

$$\overline{\alpha} = \frac{ab\omega}{2\pi} \int_{-\pi/2}^{\pi/2} \frac{1 - (\gamma_0 + \gamma_1 \sin \omega t)dt}{1 - (\gamma_0 + \gamma_1 \sin \omega t)^2 + b^2 c} \quad (25)$$

where

$$a = \omega_0 \left(\frac{\rho}{c}\right)^{1/2} \left(\frac{e_p^2}{2c\varepsilon}\right); \quad b = \frac{\omega_c}{\omega_0}; \quad c = \left(1 + \frac{\omega_0^2}{\omega_c \omega_D}\right);$$

$$\gamma_0 = \frac{\mu E_0}{V_0}; \quad \gamma_1 = \frac{\mu E_1}{V_0}$$

To integrate Eq(25) we make the following transformation $x = -\gamma_1 \sin(\omega t)$ then $dt = \dfrac{dx}{\omega\sqrt{\gamma_1^2 - x^2}}$ and Eq(25) becomes

$$\overline{\alpha} = \dfrac{ab}{2\pi}\int_{-\gamma_1}^{\gamma_1} \dfrac{(1-\gamma_0)+x}{1-(\gamma_0-x)^2 + D^2} \cdot \dfrac{dx}{\sqrt{\gamma_1^2 - x^2}} \quad (26)$$

here $D^2 = b^2 c$.

Because of the complex nature of the integral in Eq(26). We further perform another transformation i.e.

$$x' = \dfrac{x}{1-\gamma_0} \Rightarrow dx' = \dfrac{dx}{1-\gamma_0};$$

$$D' = \dfrac{D}{1-\gamma_0}; \gamma' = \dfrac{\gamma_1}{1-\gamma_0}$$

which transforms Eq(26) to the following expression

$$\overline{\alpha} = \dfrac{ab}{2\pi}\int_{-\gamma'}^{\gamma'} \dfrac{1+x'}{(1+x')^2 + D'^2} \dfrac{dx'}{\sqrt{\gamma'^2 - x'^2}} \quad (27)$$

After cumbersome manipulation we obtain for $\overline{\alpha}$ the expression

$$\overline{\alpha} = \dfrac{ab}{2}\dfrac{1}{1-\gamma_0}\left(\dfrac{1-A+(A^2-\gamma'^2)^{1/2}}{A^2-\gamma'^2}\right)^{1/2}$$

$$= \dfrac{1}{2}\dfrac{\omega C}{V_0}\dfrac{\ell^2 \rho}{2c\varepsilon}\dfrac{1}{1-\gamma_0}\left(\dfrac{1-A+(A^2-\gamma'^2)^{1/2}}{A^2-\gamma'^2}\right)^{1/2}$$

$$= \dfrac{1}{4}\dfrac{w_c}{v_0}\dfrac{k_s}{1-\gamma_0}\left(\dfrac{1-A+(A^2-\gamma'^2)^{1/2}}{A^2-\gamma'^2}\right)^{1/2} \quad (28)$$

where

$$A = \dfrac{1+D'^2+\gamma'^2}{2}$$

$$= \dfrac{1}{2}\left[1+\dfrac{1}{(1-\gamma_0)^2}\left\{\dfrac{\omega_c^2}{\omega_0^2}\left(1+\dfrac{\omega_0^2}{\omega_c\omega_D}\right)+\gamma_1^2\right\}\right]$$

**RESULTS, DISCUSSION AND CONCLUSION**

In order to calculate the attenuation and velocity of the propagating ultrasonic wave in GaN, the electromechanical coupling coefficient $x$ has to be evaluated using the parameters of Ridley [**19, 20**], O'Clock and Duffy [**21**], and Shimada et al [**22**] who have tabulated the electric and elastic constants for GaN.

The effective piezoelectric constants for the wurtzite structure for LA and TA modes are [**23**]

$$e_L = (2e_{15} + e_{31})Sin^2\theta Cos\theta + e_{33} Cos^3\theta, \quad (29)$$

$$e_T = (e_{33} - e_{15} - e_{31})Cos^2\theta Sin\theta + e_{15} Sin^3\theta \quad (30)$$

where $\theta$ is the angle between the direction of the propagation and the c-axis. The electromechanical coupling coefficient is [**19, 20**]

$$\chi_{av}^2 = \dfrac{\langle e_L^2 \rangle}{\varepsilon c_L} + \dfrac{\langle e_T^2 \rangle}{\varepsilon c_T} \quad (31)$$

where $c_L$ and $c_T$ are the angular averages off elastic constants describing the propagation of *LA* and *TA* waves, respectively [**23**] and $\langle e_L^2 \rangle, \langle e_T^2 \rangle$ are the spherical averages of piezoelectric constants $e_L$ and $e_T$ respectively, which are

$$\langle e_L^2 \rangle = \dfrac{1}{7}e_{33}^2 + \dfrac{4}{35}e_{33}(e_{31}+2e_{15}) + \dfrac{8}{105}(e_{31}+2e_{15})^2 \quad (32)$$

$$\langle e_T^2 \rangle = \frac{2}{35}(e_{33} - e_{15} - e_{31})^2$$
$$+ \frac{16}{105} e_{15}(e_{33} - e_{15} - e_{31}) + \frac{16}{35} e_{15}^2 \qquad (33)$$

We analysis the result of eq(27) for bulk GaN in a constant electric field and slowly changing a.c electric field using the value of $\chi$ in [28]. Firstly the case where the a.c electric field is absent $\gamma_1 = 0$, it is observed that as the constant electric field increases the peak of the acoustoelectric gain also increase. This is due to the fact that the mobility of the electron increases with the increase in the constant electric field (ref Eq (15)) see fig. 1.

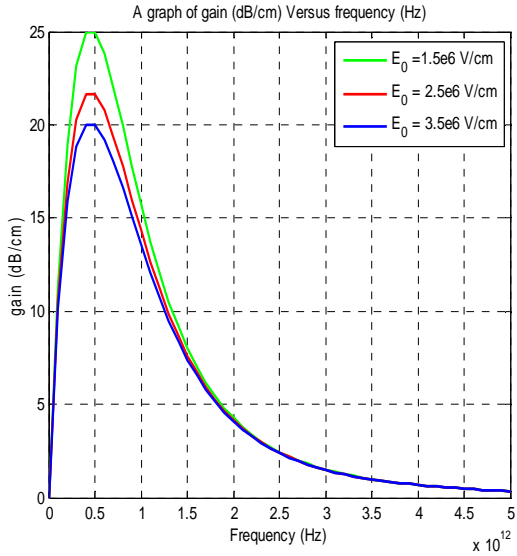

Fig. 1.

This result has also been obtained in [28]. Secondly, the case when the a.c field is applied to the system, and d.c is switched off is also considered in fig.2.

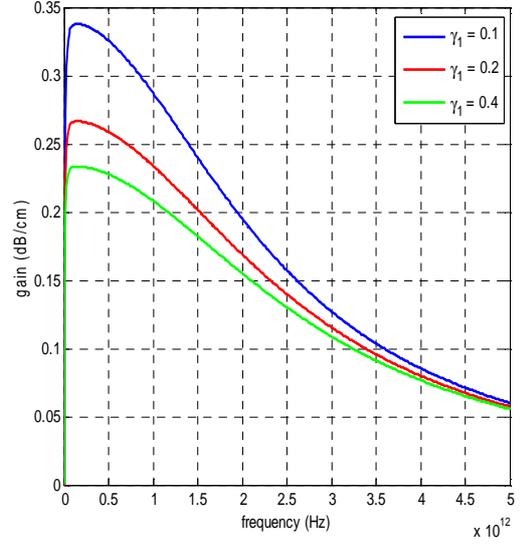

Fig. 2: The plot of acoustoelectric gain against frequency.

It is observed that as the amplitude of the a.c field increases the peaks of the gain decreases. Like the case of the d.c field, the peaks occur at specific frequencies. Thirdly, the situation where both the d.c and a.c field are present is considered (see fig. 3).

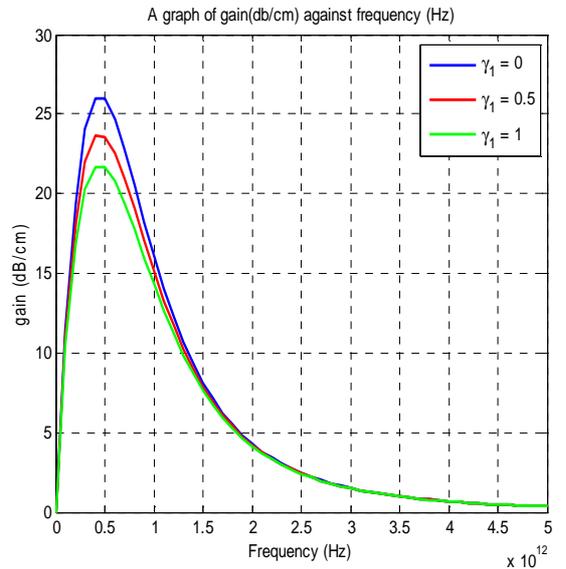

Fig . 3:

It is interesting to note that when the acoustoelectric gain is plotted against the frequency of the acoustic wave for given values of the a.c electric field $(\gamma_1)$ the maximum gain drops as $\gamma_1$ increases. This is because in the presence of a.c electric field electron bunching is slowed down and hence interaction with acoustic wave is reduced. The a.c electric field is therefore modulating the direct current.

Another excited result is the case when the d.c electric field is kept at about $0.4 \times 10^3$ V/cm and the acoustoelectric gain is plotted against the acoustic wave frequency for a given a.c electric field (see fig. 4).

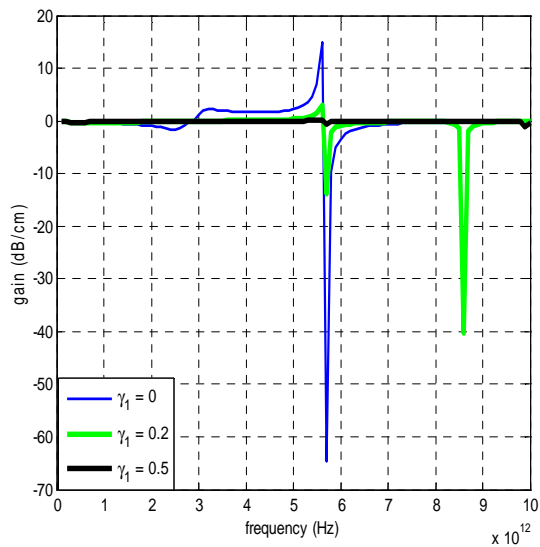

Fig. 4: Acoustoelectric gain against frequency when $E_0 = 4*10e2$ V/cm.

It is observed that the sample becomes transparent over wide range of frequency then suddenly there is huge amplification at around $5.8 \times 10^{12}$ Hz. What is intriguing is that as the a.c electric field increases a second resonant peak is observed at around $8.7 \times 10^{12}$ Hz. with further increase of the a.c field, the gain becomes zero and the sample becomes completely transparent. It is our opinion that the sample can be used as maser in this region of frequencies.

In conclusion we have studies the attenuation (amplification) of acoustic wave in GaN under a direct current and a slowly changing a.c electric field and suggest the use of this material as maser.